\documentclass[12pt]{article}
\usepackage{epsf}
\usepackage{epsfig}
\usepackage{amsmath}

\def\beq{\begin{equation}}
\def\eeq{\end{equation}}

\def\slash#1{#1 \hskip-0.45em /}

\def\eps{\varepsilon}
\def\sh{\hat s}
\def\CR{\nonumber\\}
\def\noi{\noindent}

\allowdisplaybreaks[2]

\begin{document}

\begin{titlepage}
\begin{flushright}
 PITHA 04/07\\
 hep-ph/0403185\\
 March 2004
\end{flushright}
\vskip 2.4cm

\begin{center}
\boldmath
{\Large\bf Analytic two--loop virtual corrections to $b\to d\ell^+\ell^-$}
\unboldmath 
\vskip 2.2cm
{\sc Dirk~Seidel}
\vskip .5cm
{\it Institut f\"ur Theoretische Physik~E, RWTH Aachen, 
52056 Aachen, Germany}

\vskip 2.5cm

\end{center}

\begin{abstract}
\noi
We compute the two--loop virtual corrections to the flavour changing neutral current process $b\to d\ell^+\ell^-$. As calculation techniques integration by parts identities and the method of differential equations are used. The result is presented in closed form as a function of $q^2/m_b^2$, where $q^2$ is the invariant mass of the lepton pair and $m_b$ is the $b$--quark mass.
\end{abstract}

\vskip 2.5cm

\vfill

\end{titlepage}


\section{Introduction}

Flavour changing radiative $B$--decays play a crucial role in testing the standard model, because they proceed entirely through loops and are therefore sensitive to the propagation of virtual new particles. At present, one of the most stringent constraints on new physics from rare $B$--decays arise from the inclusive decays $B\to X_s\gamma$ and $B\to X_s\ell^+\ell^-$ which have been studied in great detail in the literature~\cite{bsg,bsgll,bsll}. Within the present uncertainties, the experimental results~\cite{Kaneko:2002mr,Aubert:2003rv,Nakao:2003gc} for both decays are fully compatible with the theoretical predictions~\cite{Buras:2002tp,Ghinculov:2003qd} arising in the SM. 

Concerning the two--body decay $B\to X_s\gamma$ calculations are complete at NLO in QCD~\cite{Buras:2002tp}. For the process $B\to X_s\ell^+\ell^-$ semi--numerical results are available for the low and high range of $q^2$ (invariant mass of the lepton pair), i.e. after cutting out the nonperturbative $\bar u u$ and $\bar c c$ resonances~\cite{Ghinculov:2003qd}. There only the numerically small Wilson coefficients arising from penguin contributions were partly neglected. In these two decay modes, the branching ratio and the invariant mass distribution/forward--backward--asymmetry for the latter one are important observables that can be used to put constraints on parameters of models beyond the SM. 
 
The transitions $b\to s\gamma$ and $b\to s\ell^+\ell^-$ have, to a good approximation, vanishing CP--asymmetry, because the CKM--element $V_{us}^* V_{ub}$ can safely be neglected and therefore the CP--violating phase drops out of all observables. This changes when one deals with the $b\to d$ transition, because $V_{ud}^* V_{ub}$ is of the same order than the other two matrix elements contributing to the process. Therefore we have, in addition to the before mentioned observables another one, the CP--asymmetry which can be used to test the SM or to constrain the angles of the CKM--triangle. 

As a consequence of the new CKM--structure, at NLO one has to compute in addition to the two--loop diagrams with an internal $c$--quark propergating~\cite{Asatryan:2001zw}, the ones where the $c$--quark is replaced by a $u$-quark. We present the non--factorizable two--loop virtual corrections in an analytic form for arbitrary $q^2$, where $q^2$ is the invariant mass of the lepton pair.

For the decay mode with a real photon in the final state, the results can completely be extracted from~\cite{Buras:2002tp}. In that paper the diagrams were calculated for a whole range of operators in and beyond the standard model. Here we restrict ourselves to SM operators and furthermore we neglect the penguin Wilson coefficients because of their smallness. Recently Asatrian et.al.~\cite{Asatrian:2003vq} presented the results for $B\to X_d\ell^+\ell^-$ where the same non--factorizable two--loop corrections were calculated. They have presented it as a power series in $q^2/m_b^2$, as we give analytic formulae for the corresponding diagrams. After expanding the results given in this paper, we fully agree with their findings.

The phenomenological analysis of the decay will be done in~\cite{us}, so we restrict ourselves here to the pure computation of the two--loop diagrams.

The paper is organized as follows. In the next section the effective Hamiltonian for the considered process is given. Section~\ref{SecRedMI} contains the description of the method used to reduce the two--loop integrals to a few scalar Master Integrals~(MIs). The following section deals with the calculation of these integrals. Then we give the results for the virtual two--loop corrections and finally some concluding remarks are given. In the appendices the results for the MIs, the unrenormalized virtual corrections and the one--loop integrals needed for renormalization are given.

\section{The Effective Hamiltonian}

\label{SecEffHam}

In the Standard Model the weak effective Hamiltonian for $b\to d$ transitions is given by
\beq
H_{\rm eff} = \frac{4G_F}{\sqrt{2}}
\left[ \sum_{q=u,c}
\lambda_{q}^{(d)}(C_1\, {\cal O}_1^q+C_2\, {\cal O}_2^q)
-\lambda_{t}^{(d)}\sum_{i=3}^{10}C_i\,{\cal O}_i \right]\, + \mbox{h.c.},
\eeq
where $\lambda_q^{(d)}=V_{qd}^* V_{qb}$.

In order to match our computation with the NNLL results for the $C_i$, we use the operator basis~\cite{Chetyrkin:1997gb}
\begin{xalignat}{2}
\label{opBasis}
{\cal O}_1^q&=(\bar d_L\, \gamma_\mu T^a \,q_L)
(\bar q_L\, \gamma^\mu T^a \,b_L),&
{\cal O}_2^q&=(\bar d_L\, \gamma_\mu \,q_L)
(\bar q_L\, \gamma^\mu \,b_L),\CR
{\cal O}_3&=(\bar d_L\, \gamma_\mu \,b_L)
\sum\hspace{-.5mm}{}_q\, (\bar q\, \gamma^\mu q),&
{\cal O}_4&=(\bar d_L\, \gamma_\mu T^a \,b_L)
\sum\hspace{-.5mm}{}_q\, (\bar q\, \gamma^\mu \,T^a \,q),\CR
{\cal O}_5&=(\bar d_L\, \gamma_\mu\gamma_\nu\gamma_\rho \,b_L)
\sum\hspace{-.5mm}{}_q\, (\bar q\, \gamma^\mu\gamma^\nu\gamma^\rho \,q),&
{\cal O}_6&=(\bar d_L\, \gamma_\mu\gamma_\nu\gamma_\rho T^a \,b_L)
\sum\hspace{-.5mm}{}_q\, (\bar q\, \gamma^\mu\gamma^\nu\gamma^\rho\,T^a \,q),\CR
{\cal O}_7&=\frac{g_{\rm em}m_b}{g_s^2}\,
(\bar d_L\, \sigma^{\mu\nu}b_R)\,F_{\mu\nu},&
{\cal O}_8&=\frac{m_b}{g_s}\,
(\bar d_L\, \sigma^{\mu\nu}T^a b_R)\,G_{\mu\nu}^a,\CR
{\cal O}_9&=\frac{g_{\rm em}^2}{g_s^2}\,
(\bar d_L \,\gamma_\mu\,b_L)
\sum\hspace{-.5mm}{}_\ell\, (\bar\ell \, \gamma^\mu \, \ell),&
{\cal O}_{10}&=\frac{g_{\rm em}^2}{g_s^2}\,
(\bar d_L \,\gamma_\mu\,b_L)
\sum\hspace{-.5mm}{}_\ell\, (\bar\ell \, \gamma^\mu\gamma_5 \, \ell),
\end{xalignat}
where the sum over $q$ and $\ell$ extends over all light quark and lepton fields, respectively. $g_{\rm em}(g_s)$ is the electromagnetic (strong) coupling constant, $q_L$ and $q_R$ are the left and right chiral quark fields, $F_{\mu\nu}(G_{\mu\nu})$ is the electromagnetic (gluonic) field strength tensor and $T^a$ are the color matrices.

The one-- and two--loop matching conditions for the Wilson coefficients at the scale $\mu_W\sim M_W$
\beq
C_i(\mu_W)=C_i^{(0)}(\mu_W)+\frac{\alpha_s}{4\pi}\,C_i^{(1)}(\mu_W)
+\left(\frac{\alpha_s}{4\pi}\right)^2\,C_i^{(2)}(\mu_W)+O(\alpha_s^3)
\eeq
can be found in \cite{Bobeth:1999mk}, whereas the anomalous dimension matrix
\beq
\gamma=\frac{\alpha_s}{4\pi}\,\gamma^{(0)}
+\left(\frac{\alpha_s}{4\pi}\right)^2\gamma^{(1)}
+\left(\frac{\alpha_s}{4\pi}\right)^3\gamma^{(2)}+O(\alpha_s^4)
\eeq
which is needed to evaluate the coefficients down to the scale $\mu\sim m_b$ has been calculated to NNLL order in \cite{Gambino:2003zm}.

It has been shown that decay amplitudes for the decay of a $B$--meson into light mesons in the kinematic region of large recoil of the light meson can be systematically computed in terms of form factors, light--cone distribution amplitudes and perturbative hard scattering kernels~\cite{Beneke:1999br,Beneke:2000ry,Beneke:2000wa}. Schematically, the amplitude can be represented as~\cite{Beneke:2001at} 
\begin{equation}
\langle \ell^+\ell^- \rho_a|H_{\rm eff}|{B}\rangle
= C_a \,\xi_a + \Phi_B\otimes T_a\otimes \Phi_\rho,
\end{equation}
where $a=\perp,\parallel$ refers to a transversely and longitudinally polarized $\rho$--meson, respectively. In this equation $\xi_a$ represent universal heavy--to--light form factors~\cite{Beneke:2000wa,Charles:1998dr} and $\Phi$ light--cone--distribution amplitudes. The factors $C_a$ and $T_a$ are calculable in renormalization--group improved perturbation theory. In this paper we compute the NLO non--factorizable two--loop contributions to the $C_a$. The corresponding diagrams are shown in fig.~\ref{AbballeDiag}. In addition, there are factorizable two--loop vertex corrections. As the difficult part of the calculation lies in the non--factorizable diagrams , we refrain from presenting the factorizable ones. For the exclusive decay these have been calculated in~\cite{Beneke:2000wa} whereas we refer to~\cite{Asatrian:2003vq} concerning the inclusive mode.

\section{Reduction to Master Integrals}
\label{SecRedMI}

\begin{figure}[t]
\begin{center}
\psfig{file=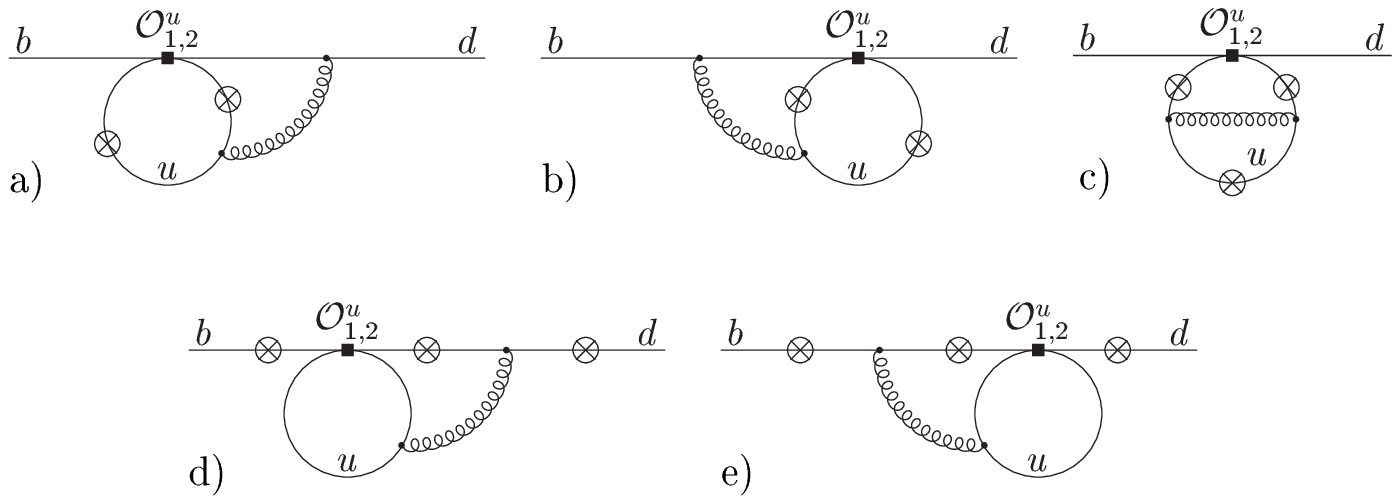,width = 15.8cm}
\vspace{.1cm}
\caption{\label{AbballeDiag} Two--loop contributions to $\langle\gamma^* d|H_{\rm eff}|b\rangle$ that have been calculated in this paper. The circled crosses mark the possible insertations of the virtual photon line.}
\end{center}
\end{figure}

In this section we present the calculation of the two--loop matrix elements 
$\langle d\gamma^*|{\cal O}_{1,2}^u|b\rangle$, were $\gamma^*$ denotes a virtual photon. From these we easily get the quantities 
$\langle d\ell^+\ell^-|{\cal O}_{1,2}^u|b\rangle$ we finally wish to compute. The corresponding diagrams are shown in fig.\ref{AbballeDiag}.

It is always possible to decompose the matrix element into the following form:
\begin{eqnarray}
\langle\gamma^*(q,\mu) d(p')|H_{\rm eff}|b(p)\rangle &\equiv&
\langle d(p')|\bar d\, X^\mu\, b|b(p)\rangle=
F^{(q)}(q^2)\langle d(p')|\bar d(1+\gamma^5)q^\mu\,b|b(p)\rangle\CR
&&+F^{(7)}(q^2)\langle d(p')|\bar d(1+\gamma^5)\sigma^{\mu\nu}q_\nu
 \, b|b(p)\rangle\CR
&&+F^{(9)}(q^2)\langle d(p')|\bar d(1+\gamma^5)(q^2\gamma^\mu-q^\mu\slash q)b|b(p)\rangle
\end{eqnarray}
The scalar form factors $F^{(q)}, F^{(7)}, F^{(9)}$ are obtained by taking the trace
\begin{eqnarray}
F^{(i)}(q^2)={\rm Tr}\left(P_i^\mu X_\mu\right)
\end{eqnarray}
with the projectors
\begin{eqnarray}
P_i^\mu=(\,\slash p+m_b)\left(C_{i1}\,q^\mu+C_{i2}\,p^\mu+C_{i3}\,\gamma^\mu\right)\slash p'
\end{eqnarray}
and some scalars $C_{ij}$. The factors $(\,\slash p+m_b)$ and \,$\slash p'$ account for the onshell condition of the $b-$ and $s-$quark lines, respectively. $F^{(q)}$ can give a nonzero contribution to individual diagrams but vanishes in the sum of the diagrams because of electromagnetic gauge invariance. This gives a good check of the computation which is described in the following. After projecting out the form factors, these are of the form
\begin{eqnarray}
\label{prop}
F^{(i)}(q^2)=\sum_j c_{ij} \int d^d k\,d^d l\,
 \frac{{\cal S}_1^{n_{1j}}{\cal S}_2^{n_{2j}}\dots {\cal S}_7^{n_{7j}}}
 {{\cal P}_1^{m_{1j}}{\cal P}_2^{m_{2j}}\dots {\cal P}_9^{m_{9j}}},
\end{eqnarray}
where the following scalar products and propagators containing loop momenta appear:
\begin{alignat}{4}
{\cal S}_1&=k^2,&\qquad {\cal S}_2&=l^2,&\qquad {\cal S}_3&=k\cdot l,
 &\qquad {\cal S}_4&=k\cdot p,\CR
 {\cal S}_5&=k\cdot q,& {\cal S}_6&=l\cdot p,&\qquad {\cal S}_7&=l\cdot q,
\end{alignat}
\begin{alignat}{3}
 {\cal P}_1&=\frac{1}{k^2},&\quad {\cal P}_2&=\frac{1}{(k+p-q)^2},&
\quad {\cal P}_3&=\frac{1}{(l+q)^2},\CR
 {\cal P}_4&=\frac{1}{l^2},&\quad {\cal P}_5&=\frac{1}{(l+k+q)^2},&
\quad {\cal P}_6&=\frac{1}{(k-p+q)^2-m_b^2},\CR
 {\cal P}_7&=\frac{1}{(l+k)^2},&\quad {\cal P}_8&=\frac{1}{(k+p)^2},&
\qquad {\cal P}_9&=\frac{1}{(k-p)^2-m_b^2}.
\end{alignat}
For the diagrams in fig.~\ref{AbballeDiag} we get at most five powers of the propagators ${\cal P}_i$  and three powers of the scalar products ${\cal S}_i$. 

The next step of the calculation is to reduce the above integrals (in our case there are a few hundred) to a few MIs by means of the so--called Laporta algorithm~\cite{Laporta:1996mq,Laporta:2001dd} using Integration by Part~(IBP)~\cite{Tkachov:wb,Chetyrkin:qh} and Lorentz invariance~\cite{Gehrmann:1999as} identities. IBP identities use the fact that in dimensional regularization we have the following equalities:
{\begin{eqnarray}
\int d^d k\,d^d l\, \frac{\partial}{\partial k^\mu}\left( v^\mu \,
 \frac{{\cal S}_1^{n_{1}} {\cal S}_2^{n_{2}}\dots  {\cal S}_7^{n_{7}}}
 { {\cal P}_1^{m_{1}} {\cal P}_2^{m_{2}}\dots  {\cal P}_9^{m_{9}}} \right)&=&0,\CR
\int d^d k\,d^d l\, \frac{\partial}{\partial l^\mu}\left( v^\mu \,
 \frac{ {\cal S}_1^{n_{1}} {\cal S}_2^{n_{2}}\dots  {\cal S}_7^{n_{7}}}
 { {\cal P}_1^{m_{1}} {\cal P}_2^{m_{2}}\dots  {\cal P}_9^{m_{9}}} \right)&=&0
\end{eqnarray}
where
$v^\mu=k^\mu,\, l^\mu,\, p^\mu,\, q^\mu$. For the kinematics in the present process the Lorentz invariance identity for an arbitrary scalar integral takes the form
\begin{eqnarray}
\hspace{-.5cm}
\int d^d k\,d^d l\,
\left[p \cdot q\, p^\mu\,\frac{\partial}{\partial p^\mu}
-p \cdot q\, q^\mu\,\frac{\partial}{\partial q^\mu}
+q^2\, p^\mu\,\frac{\partial}{\partial q^\mu}
-p^2\, q^\mu\,\frac{\partial}{\partial p^\mu}\right]
\frac{ {\cal S}_1^{n_{1}} {\cal S}_2^{n_{2}}\dots  {\cal S}_7^{n_{7}}}
 { {\cal P}_1^{m_{1}} {\cal P}_2^{m_{2}}\dots  {\cal P}_9^{m_{9}}}=0.
\end{eqnarray}
This way we get nine identities for each scalar integral. These identities will contain other scalar integrals with different numbers of scalar products and propagators. After making up the identities for all the appearing integrals, the aim is to solve these linear equations for integrals of a simpler type. In principle this is a simple task but with the diagrams considered here we get linear equation systems up to about 3200 equations and therefore some organization is required to solve them. 

As stated above, we start with integrals with at most five powers of different propagators. It is possible to express all of these in terms of integrals containing four different propagators. Concerning these simpler topologies, there exist some integrals where no equation relating these to simpler topologies can be found by that method. The same happens for integrals with three different propagators. The remaining integrals are the MIs which we solve by the method of differential equations. Fig.~\ref{AbbMI} lists all the MIs appearing.

\section{Calculation of the Master Integrals}

\begin{figure}[t]
\begin{center}
\begin{picture}(100,80)(0,-25)
  \psfig{file=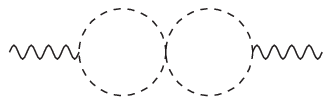,width = 3.5cm}
  \put(-50,-15){a)}
\end{picture}
\begin{picture}(100,80)(0,-25)
  \psfig{file=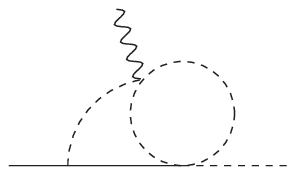,width = 3.5cm}
  \put(-50,-15){b)}
\end{picture}
\begin{picture}(100,80)(0,-25)
  \psfig{file=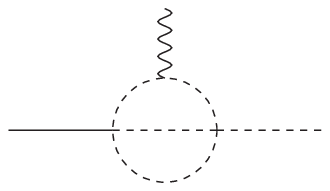,width = 3.5cm}
  \put(-50,-15){c)}
\end{picture}
\begin{picture}(100,80)(0,-25)
  \psfig{file=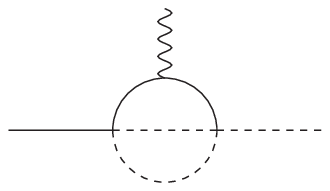,width = 3.5cm}
  \put(-50,-15){d)}
\end{picture}
\begin{picture}(100,80)(0,-25)
  \psfig{file=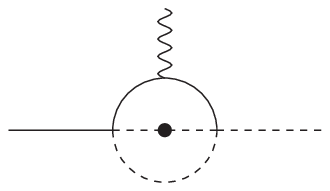,width = 3.5cm}
  \put(-50,-15){e)}
\end{picture}
\begin{picture}(100,80)(0,-25)
  \psfig{file=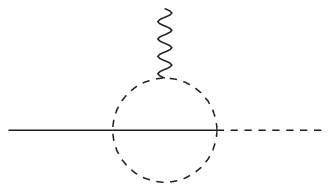,width = 3.5cm}
  \put(-50,-15){f)}
\end{picture}
\begin{picture}(100,80)(0,-25)
  \psfig{file=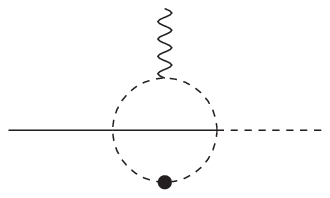,width = 3.5cm}
  \put(-50,-15){g)}
\end{picture}
\begin{picture}(100,80)(0,-25)
  \psfig{file=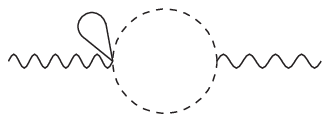,width = 3.5cm}
  \put(-50,-15){h)}
\end{picture}
\begin{picture}(100,60)(0,-25)
  \psfig{file=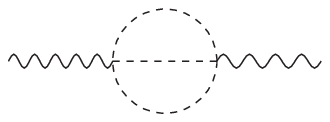,width = 3.5cm}
  \put(-50,-15){i)}
\end{picture}
\begin{picture}(100,60)(0,-25)
  \psfig{file=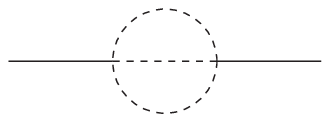,width = 3.5cm}
  \put(-50,-15){j)}
\end{picture}
\begin{picture}(100,60)(0,-25)
  \psfig{file=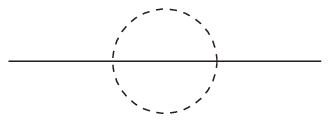,width = 3.5cm}
  \put(-50,-15){k)}
\end{picture}
\begin{picture}(100,60)(0,-25)
  \psfig{file=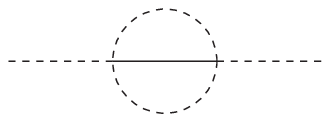,width = 3.5cm}
  \put(-50,-15){l)}
\end{picture}
\caption{\label{AbbMI}
 Scalar Master Integrals appearing in our calculations. The dashed lines denote massless propagators whereas the solid lines represent the ones with mass $m_b$. Dotted propagators have to be squared. Wavy/solid/dashed external lines are onshell by the amount $q^2/m_b^2/0$, respectively.}
\end{center}
\end{figure}

The last step in the calculation of the form factors is the evaluation of the MIs. We find five MIs with three propagators and seven with four different propagators, where in the latter case, two of the propagators show up squared. The 3--denominator topologies in fig.~\ref{AbbMI}h-l and the MI in fig.~\ref{AbbMI}a which factorizes into two one--loop integrals can be calculated trivially. Nevertheless we have given the results for completeness in appendix~\ref{appMI}, like the results for the other MIs. For the remaining MIs we use the method of differential equations~\cite{Kotikov:1990kg,Remiddi:1997ny,Bonciani:2003te}. In our case, the equations take the form
\begin{eqnarray}
\label{DiffEq}
\frac{\partial}{\partial q^2}\,{\rm MI}(q^2,m_b^2)=\frac{1}{m_b^2-q^2}\,q^\mu
\left(\frac{\partial}{\partial p^\mu}-\frac{\partial}{\partial q^\mu}
\right) {\rm MI}(q^2,m_b^2)
\end{eqnarray}
where the derivative is understood to act on the integrand of the corresponding MI. After evaluating the right hand side of (\ref{DiffEq}), there will be terms containing the MI itself but also terms with other scalar integrals. We now use the same method as described in the last section to express these in terms of MIs. We end up with two different possibilities.

In the first case the derivative with respect to $q^2$ of the MI gives terms proportional to the MI itself and some MIs of a simpler topology, where one propagator is completely canceled, not just reduced by one power (figs.~\ref{AbbMI}b-c). In this case we are left with an inhomogeneous linear differential equation with 3--denominator topologies as inhomogeneities that we have already solved. Solving this equation is even easier by expanding the MIs in $\eps$:
\begin{eqnarray}
{\rm MI}(q^2,m_b^2)=
 \sum_{i=-2}^\infty\frac{1}{\eps^i}\,{\rm MI}^{(i)}(q^2,m_b^2)
\end{eqnarray}
After expanding the differential equation (the coefficients are $\eps$--dependant) in $\eps$, we can solve it order by order in $\eps$, the inhomogeneous parts either coming from the 3--denominator topologies or from the part of the 4--denominator MI in the $\eps$--expansion that has already been solved.

The second possibility arises when solving the MIs that are shown in figs.~\ref{AbbMI}d-e and figs.~\ref{AbbMI}f-g, respectively. After reducing (\ref{DiffEq}) to contain only MIs, we are left with a coupled inhomogeneous differential equation for the two 4--denominator topologies. When we expand this system in $\eps$, the equations decouple order by order. In both cases we used Eulers variation of the constant to solve the different parts.

This method can be used in principle to get the MIs to any desired order in $\eps$. As described in the previous section, the form factors are expressed in terms of MIs with certain coefficients, which can depend on $\eps$. Since we want to know them up to the finite parts in the limit $\eps\to 0$, it turns out that we do not have to exceed the second order in the expansion for the MIs. For one of the integrals (fig.\ref{AbbMI}e) we actually only need to go to the $1/\eps$--term.

\section{Renormalized matrix elements}
\label{secRenFF}

In this section the main result of this paper, the $\overline{{\rm MS}}$--renormalized NLO non--factorizable two--loop virtual correction relevant for the decay $b\to d \ell^+\ell^-$, is given. We present the matrix--elements as in~\cite{Asatrian:2003vq}, that is after the decomposition
\begin{eqnarray}
\langle d\ell^+\ell^-|{\cal O}_i^u|b\rangle_{\stackrel{\text{\scriptsize non-fact.}}{\text{two-loop}}}=
\frac{\alpha_s}{4\pi}\left(F_{i,u}^{(7)}\langle {\cal \tilde O}_7\rangle_{\rm tree}
+F_{i,u}^{(9)}\langle {\cal \tilde O}_9\rangle_{\rm tree}\right)
\end{eqnarray}
where $\langle {\cal \tilde O}_{7}\rangle_{\rm tree}$ and $\langle {\cal \tilde O}_{9}\rangle_{\rm tree}$ denote the tree level matrix elements of ${\cal \tilde O}_{7}$ and ${\cal \tilde O}_{9}$, respectively. Furthermore we have defined the rescaled operators 
\begin{eqnarray}
{\cal \tilde O}_{7/9}=\frac{\alpha_s}{4\pi}\,{\cal O}_{7/9}.
\end{eqnarray}
Note that our sign in the above definition differs from that in~\cite{Asatrian:2003vq}.

As we use the MS--scheme renormalization constants, all unrenormalized l--loop form factors are assumed to be multiplied by $(4\pi)^{-l\eps}e^{l\eps\gamma}$. They are given in appendix~\ref{appUnrenFF} and~\ref{appOneloop}. Since the Wilson coefficients $C_{3-6}$ for the penguin contributions were neglected, the part of the renormalized effective Hamiltonian which contains the counterterms needed for renormalization reads
\begin{eqnarray}
\delta H_{\rm eff} = \frac{4G_F}{\sqrt{2}}
\sum_{i=1}^2
\lambda_{q}^{(d)}C_i
\left[ \sum_{j=1}^2 \delta Z_{ij}{\cal O}_j^u
+\sum_{j=3}^{10} \delta Z_{ij}{\cal O}_j
+\sum_{j=11}^{12} \delta Z_{ij}{\cal O}_j^u
\right]\, + \dots.
\end{eqnarray}
The operators ${\cal O}_{1-10}$ are given in (\ref{opBasis}), whereas ${\cal O}_{11/12}$ are evanescent operators vanishing in $d=4$. In order to match our computation with the Wilson coefficients of~\cite{Gambino:2003zm}, we choose them as
\begin{eqnarray}
{\cal O}_{11}^u &=& (\bar d_L\, \gamma_\mu\gamma_\nu\gamma_\rho \,T^a\,u_L)
(\bar u_L\, \gamma^\mu\gamma^\nu\gamma^\rho \,T^a\,b_L)-16{\cal O}_1^u,\CR
{\cal O}_{12}^u &=& (\bar d_L\, \gamma_\mu\gamma_\nu\gamma_\rho \,u_L)
(\bar u_L\, \gamma^\mu\gamma^\nu\gamma^\rho \,b_L)-16{\cal O}_2^u.
\end{eqnarray}
The operator renormalization matrix $Z=1+\delta Z$ is given by~\cite{Bobeth:1999mk}
\begin{align}
\delta Z&=\frac{\alpha_s}{4\pi}\left(
\begin{array}{c c c c c c c c c c c c}
-\frac{2}{\eps} & \frac{4}{3\eps} & \,0\, & -\frac{1}{9\eps} & \,0\, &\, 0 \,&
\,0 \, & \,0\, & -\frac{16}{27\eps} &
\,0\, & \frac{5}{12\eps} & \frac{2}{9\eps}\vspace{.5cm}\\ 
\frac{6}{\eps} &\, 0 \,&\, 0\, & \frac{2}{3\eps} & \,0\, & \,0\, &
\,0\,  & \,0\, & 
-\frac{4}{9\eps} & \,0\, & \frac{1}{\eps} &\, 0\,
\end{array}
\right)\CR
&+\left(\frac{\alpha_s}{4\pi}\right)^2\left(
\begin{array}{c c c c c c c c c c c c}
\,0\, & \,0\, & \,0\, & \,0\, & \,0\, & \,0\, &
\,-\frac{58}{243\eps}\, & \,0 \,& \,
-\frac{64}{729\eps}+\frac{1168}{243\eps^2} \,&
\,0\, & \,0\, & \,0\vspace{.5cm}\\ 
\,0\, & \,0\, & \,0\, & \,0\,  & \,0\, & \,0\, &
\,\frac{116}{81\eps}\, & \,0\, & \,0
+\frac{776}{243\eps}+\frac{148}{81\eps^2}\,&
\,0\, & \,0\, & \,0
\end{array}
\right) + O(\alpha_s^3)
\end{align}
As a lot of entries in this matrix are zero, we only have to compute the one--loop matrix elements of ${\cal O}_{1,2,4,11,12}^u$ and furthermore the tree--level matrix elements of ${\cal O}_{7,9}$ are needed. Mass and wave function renormalization is a higher order effect and we do not have to include it here. The strong coupling constant in the definition of the operator ${\cal O}_{9}$ has to be renormalized where $Z_{g_s}$ is given by
\begin{eqnarray}
Z_{g_s}=1-\frac{\alpha_s}{4\pi}\frac{\beta_0}{2}\frac{1}{\eps},\qquad
\beta_0=11-\frac{2}{3}\,N_f,\qquad N_f=5.
\end{eqnarray}
Note that there is no renormalization needed for $\alpha_s$ in ${\cal O}_{7}$ because mixing of the four--quark--operators into ${\cal O}_{7}$ vanishes at one--loop and therefore the corresponding component in $\delta Z$ only starts at O($\alpha_s^2$) and the coupling constant renormalization for ${\cal O}_{7}$ is a higher order effect.

As only two different color structures arise in the different diagrams, we can split the form factors up into
\begin{eqnarray}
F_{1,u}^{(7)} &=& A(s),\\
F_{2,u}^{(7)} &=& -6 A(s),\\
F_{1,u}^{(9)} &=& B(s) + 4 C(s),\\
F_{2,u}^{(9)} &=& -6 B(s) + 3 C(s),
\end{eqnarray}
where the functions $A(s)$, $B(s)$ and $C(s)$ are given below. 

The following definitions are used in the formulae:
\begin{eqnarray}
s=q^2,\qquad \hat s=\frac{s}{m_b^2},\qquad z=\frac{4m_b^2}{s}
\end{eqnarray}
\begin{xalignat}{2}
x_1 &= \frac{1}{2}+\frac{i}{2}\,\sqrt{z-1},&
x_2 &= \frac{1}{2}-\frac{i}{2}\,\sqrt{z-1},\CR
x_3 &= \frac{1}{2}+\frac{i}{2\sqrt{z-1}},&
x_4 &= \frac{1}{2}-\frac{i}{2\sqrt{z-1}}.
\end{xalignat}
$\mu\sim m_b$ denotes the renormalization scale, $\zeta$ the Riemannian Zeta function and
\begin{eqnarray}
{\rm Li}_2(x)=-\int_0^x dt\,\frac{\ln(1-t)}{t},\qquad
\end{eqnarray}
is the Dilogarithm.

The functions $A(s), B(s), C(s)$ are as follows:
\begin{eqnarray}
A(s) &=& 
-\frac{104}{243}\,\ln(\frac{m_b^2}{\mu^2})
+\frac{4\sh}{27(1-\sh)}\,
  \Big[{\rm Li}_2(\sh)
  +\ln(\sh)\ln(1-\sh)\Big]\CR
&&\hspace{-.8cm}+\frac{1}{729(1-\sh)^2}\,
  \Big[6\sh\Big(29-47\sh\Big)\ln(\sh)
  +785-1600\sh+833\sh^2
  +6\pi i\Big(20-49\sh+47\sh^2\Big)\Big]\CR
&&\hspace{-.8cm}-\frac{2}{243(1-\sh)^3}\,
  \Big[2\sqrt{z-1}\Big(-4+9\sh-15\sh^2+4\sh^3\Big){\rm arccot}(\sqrt{z-1})
  +9\sh^3\ln^2(\sh)\CR
  &&+18\pi i \sh\Big(1-2\sh\Big)\ln(\sh)\Big]\CR
&&\hspace{-.8cm}+\frac{2\sh}{243(1-\sh)^4}\,
  \Big[36\,{\rm arccot}^2(\sqrt{z-1})
  +\pi^2\Big(-4+9\sh-9\sh^2+3\sh^3\Big)\Big]
\end{eqnarray}
\begin{eqnarray}
B(s) &=& 
\frac{8}{243\sh}\Big[
  (4-34\sh-17\pi i\sh)\ln(\frac{m_b^2}{\mu^2})
  +8\sh\ln^2(\frac{m_b^2}{\mu^2})
  +17\sh\ln(\sh)\ln(\frac{m_b^2}{\mu^2})\Big]\CR
&&+\frac{(2+\sh)\sqrt{z-1}}{729\sh}\,
  \Big[-48\ln(\frac{m_b^2}{\mu^2})\,{\rm arccot}(\sqrt{z-1})
  -18\pi\ln(z-1)
  +3i\ln^2(z-1)\CR
  &&-24i\,{\rm Li}_2(-x_2/x_1)
  -5\pi^2 i
  +6i\Big(-9\ln^2(x_1)+\ln^2(x_2)-2\ln^2(x_4)\CR
    &&\hspace{1cm}+6\ln(x_1)\ln(x_2)-4\ln(x_1)\ln(x_3)
    +8\ln(x_1)\ln(x_4)\Big)\CR
  &&-12\pi\Big(2\ln(x_1)+\ln(x_3)+\ln(x_4)\Big)\Big]\CR
&&\hspace{-.8cm}-\frac{2}{243\sh(1-\sh)}\,
  \Big[4\sh\Big(-8+17\sh\Big)\Big({\rm Li}_2(\sh)+\ln(\sh)\ln(1-\sh)\Big)\CR
  &&+3\Big(2+\sh\Big)\Big(3-\sh\Big)\ln^2(x_2/x_1)
  +12\pi\Big(-6-\sh+\sh^2\Big){\rm arccot}(\sqrt{z-1})\Big]\CR
&&\hspace{-.8cm}+\frac{2}{2187\sh(1-\sh)^2}\,
  \Big[-18\sh\Big(120-211\sh+73\sh^2\Big)\ln(\sh)\CR
  &&-288-8\sh+934\sh^2-692\sh^3
  +18\pi i\sh\Big(82-173\sh+73\sh^2\Big)\Big]\CR
&&\hspace{-.8cm}-\frac{4}{243\sh(1-\sh)^3}\,
  \Big[-2\sqrt{z-1}\Big(4-3\sh-18\sh^2+16\sh^3-5\sh^4\Big){\rm arccot}(\sqrt{z-1})\CR
  &&-9\sh^3\ln^2(\sh)
  +2\pi i\sh\Big(8-33\sh+51\sh^2-17\sh^3\Big)\ln(\sh)\Big]\CR
&&\hspace{-.8cm}+\frac{2}{729\sh(1-\sh)^4}\,
  \Big[72\Big(3-8\sh+2\sh^2\Big){\rm arccot}^2(\sqrt{z-1})\CR
  &&-\pi^2\Big(54-53\sh-286\sh^2+612\sh^3-446\sh^4+113\sh^5\Big)\Big]
\end{eqnarray}
\begin{eqnarray}
C(s) &=& -\frac{16}{81}\,\ln(\frac{s}{\mu^2})
+\frac{428}{243}-\frac{64}{27}\,\zeta(3)+\frac{16}{81}\,\pi i
\end{eqnarray}

\begin{figure}[t!]
\begin{center}
  \psfig{file=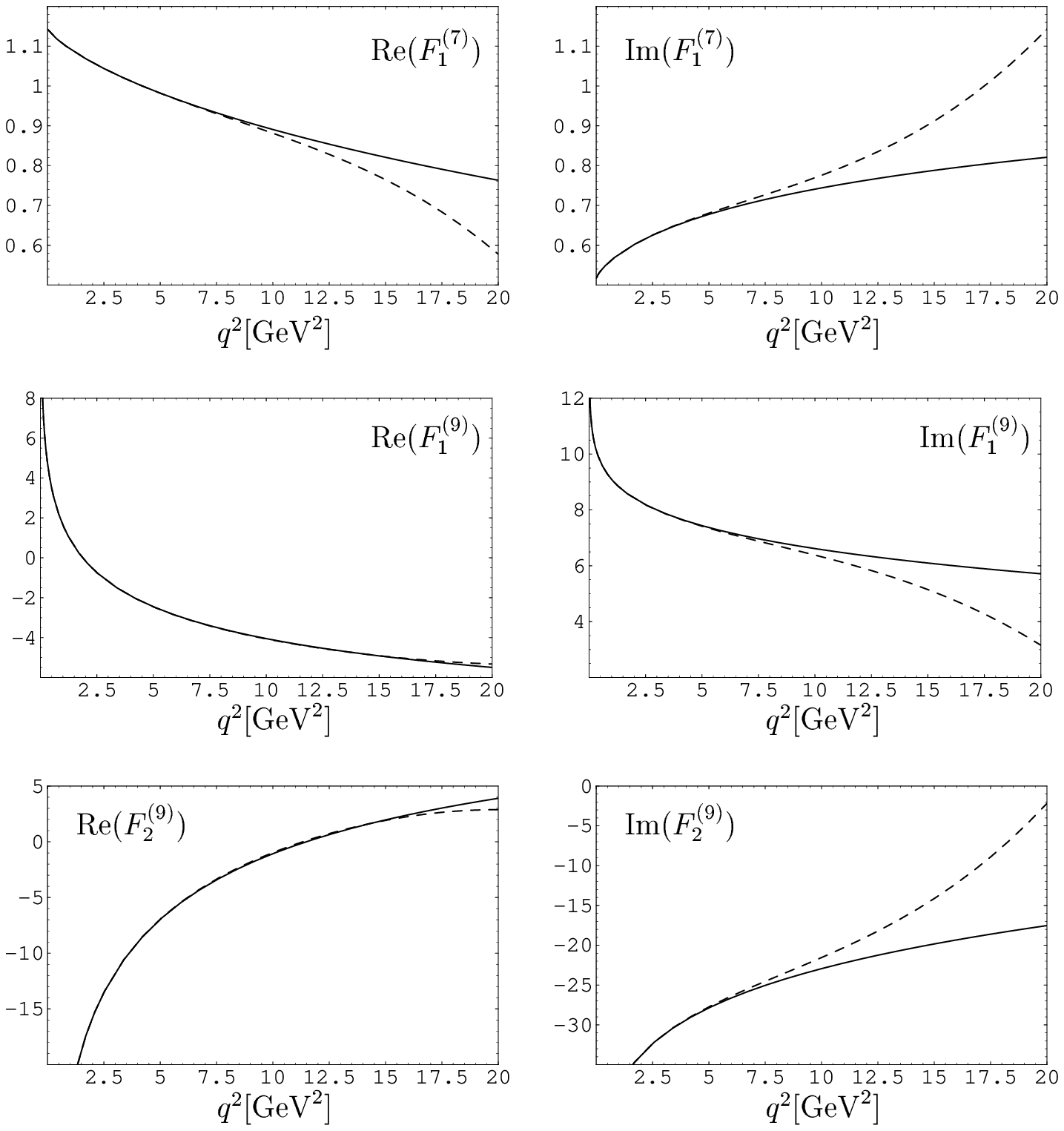,width = 15.8cm}
\caption{\label{AbbVergl}
Real and imaginary parts of the form factors $F_1^{(7)}, F_1^{(9)}, F_2^{(9)}$. The solid line represents the exact result to O($\alpha_s$), whereas the dashed line shows the result expanded in $q^2/m_b^2$ up to the third order.}
\end{center}
\end{figure}

In fig.~\ref{AbbVergl} we give a comparison between the numerical values of the NLO corrections to the form factors $F_1^{(7)}, F_1^{(9)}, F_2^{(9)}$ as computed in this paper and the result of the expansion to third order in $q^2/m_b^2$ which agrees with the result presented in~\cite{Asatrian:2003vq}. The physical ``threshold'' of the charmonium resonances begins at $M_{J/\psi}^2\approx 9.6\,\mbox{GeV}^2$, but the perturbative approximation is expected to fail earlier. Model-dependent studies of the charmonium contributions suggest that the perturbative approximation should be valid up to $q^2\approx (6-7)\,\mbox{GeV}^2$ \cite{Lim:1988yu,Kruger:1996cv}. As can be seen in fig.~\ref{AbbVergl}, the approximation up to that value of $q^2$ is fairly good but fails as expected in the high $q^2$ region, beyond the resonance. The largest deviation (at $q^2=7\,{\rm GeV}^2$) comes from ${\rm Im}(F_1^{(7)})$ and amounts to $1.3$\%.

\section{Concluding discussion}

In this paper we have presented the non--factorizable two--loop virtual corrections contributing to the process $b\to d\ell^+\ell^-$. We used the technique of integration by parts for the reduction of the few hundred integrals to a few Master Integrals. To solve these MIs we apply the method of differential equations which allowed us to calculate the two-loop diagrams analytically. Asatrian et. al.~\cite{Asatrian:2003vq} presented these calculations as an expansion in $q^2/m_b^2$, whereas our result is valid for the whole range of $q^2$. When expanded in $q^2/m_b^2$, our result agrees with the calculation made by Asatrian et. al.~\cite{Asatrian:2003vq}.

\subsection*{Acknowledgements}

I would like to thank Martin Beneke and Thomas Gehrmann for useful discussions. Furthermore I am grateful to Martin Beneke for many useful comments and for his careful reading of the manuscript. 

\begin{appendix}

\section{Solutions for the Master Integrals}
\label{appMI}

In this appendix we give the solutions for the three-- and four--point MIs needed for the calculation of the form factors for the two--loop integrals. Especially the three--point integrals and the four--point integral shown in fig.~\ref{AbbMI}a are easy to calculate but are listed here for completeness.

In the following the integrals are normalized according to~\cite{Bonciani:2003te}:
\begin{eqnarray}
\{d^d k\}=\frac{d^d k}{\pi^{d/2}\Gamma(3-d/2)}
\end{eqnarray}
We give the results up to the required order needed for our evaluation of the form factors.

\subsection{3--denominator topologies}

\begin{align}
\begin{picture}(90,30)(0,14)
 \psfig{file=MI31.ps,width = 3cm}
\end{picture} 
&=\mu^{2(4-d)}\int\{d^dk\}\{d^dl\}
  \frac{1}{{\cal P}_2{\cal P}_7{\cal P}_9}\CR
&=-m_b^2\left(\frac{m_b^2}{\mu^2}\right)^{-2\eps}\bigg\{
\frac{1}{\eps^2}
+\frac{1}{\eps}\Big(
  3-\ln(\sh)+\pi i\Big)\CR
  &\hspace{-2cm}+7-\frac{2}{3}\pi^2-3\ln(\sh)
  +\frac{1}{2}\ln^2(\sh)
  +3\pi i-\pi i \ln(\sh)+O(\eps)\bigg\}
\end{align}
\begin{align}
\begin{picture}(90,30)(0,14)
 \psfig{file=MI32.ps,width = 3cm}
\end{picture} 
&=\mu^{2(4-d)}\int\{d^dk\}\{d^dl\}
\frac{1}{{\cal P}_1{\cal P}_6{\cal P}_7}\CR
&\hspace{-2cm}=
s\left(\frac{s}{\mu^2}\right)^{-2\eps}\bigg\{
\frac{1}{4\eps}
+\frac{13}{8}+\frac{1}{2}\pi i
 +\frac{\eps}{48}\Big(
 345-28\pi^2+156\pi i\Big)\CR
&\hspace{-1.5cm}+\frac{\eps^2}{96}\Big(
 2595-364\pi^2+1380\pi i-48\pi^3 i
 -240\zeta(3)\Big)
 +O(\eps^3)\bigg\}
\end{align}
\begin{align}
\begin{picture}(90,30)(0,14)
 \psfig{file=MI33.ps,width = 3cm}
\end{picture} 
&=\mu^{2(4-d)}\int\{d^dk\}\{d^dl\}
\frac{1}{{\cal P}_2{\cal P}_6{\cal P}_9}\CR
&=
-m_b^2\left(\frac{m_b^2}{\mu^2}\right)^{-2\eps}\bigg\{
\frac{1}{2\eps^2}+\frac{5}{4\eps}+\frac{11}{8}+\frac{1}{3}\pi^2
+O(\eps)\bigg\}
\end{align}
\begin{align}
\begin{picture}(90,30)(0,14)
 \psfig{file=MI34.ps,width = 3cm}
\end{picture} 
&=\mu^{2(4-d)}\int\{d^dk\}\{d^dl\}
\frac{1}{{\cal P}_2{\cal P}_5{\cal P}_6}\CR
&\hspace{-2cm}=
-m_b^2\left(\frac{m_b^2}{\mu^2}\right)^{-2\eps}\bigg\{
\frac{1}{2\eps^2}+\frac{3}{2\eps}+\frac{7}{2}+\frac{1}{6}\pi^2
+\frac{\eps}{2}\Big(15+\pi^2-2\zeta(3)\Big)
+O(\eps^2)\bigg\}
\end{align}
\begin{align}
\begin{picture}(90,30)(0,14)
 \psfig{file=MI35.ps,width = 3cm}
\end{picture} 
&=\mu^{2(4-d)}\int\{d^dk\}\{d^dl\}
\frac{1}{{\cal P}_2{\cal P}_6{\cal P}_8}\CR
&=
\begin{picture}(90,30)(0,14)
 \psfig{file=MI32.ps,width = 3cm}
\end{picture} 
(s\rightarrow m_b^2)
\end{align}

\subsection{4--denominator topologies}

\begin{align}
\begin{picture}(90,30)(0,14)
 \psfig{file=MI41.ps,width = 3cm}
\end{picture} 
&=\mu^{2(4-d)}\int\{d^dk\}\{d^dl\}
\frac{1}{{\cal P}_2{\cal P}_4{\cal P}_6{\cal P}_7}\CR
&=
-\left(\frac{s}{\mu^2}\right)^{-2\eps}\bigg\{
\frac{1}{\eps^2}+\frac{2}{\eps}(2+\pi i)
+12-\frac{7}{3}\pi^2+8\pi i\CR
&\hspace{3cm}+2\eps\left(16-\frac{14}{3}\pi^2-2\zeta(3)
+12\pi i-\pi^3 i\right)
+O(\eps^2)\bigg\}
\end{align}
\begin{align}
\begin{picture}(90,30)(0,14)
 \psfig{file=MI42.ps,width = 3cm}
\end{picture} 
&=\mu^{2(4-d)}\int\{d^dk\}\{d^dl\}
\frac{1}{{\cal P}_1{\cal P}_6{\cal P}_7{\cal P}_9}\CR
&\hspace{-1cm}=
-\left(\frac{m_b^2}{\mu^2}\right)^{-2\eps}\bigg\{
\frac{1}{2\eps^2}
+\frac{5}{2\eps}
+\frac{57}{6}+\frac{1}{6}\pi^2
+\frac{\sh}{2(1-\sh)}\Big(
  \ln^2(\sh)-2\pi i\ln(\sh)\Big)\CR
&\hspace{-2.5cm}+\frac{\eps}{6}\Big(195+5\pi^2\Big)
-\frac{\eps}{6(1-\sh)}\Big[
  12\sh\,{\rm Li}_3(\sh)
  -12\sh\Big(\ln(\sh)-\pi i\Big){\rm Li}_2(\sh)\CR
  &\hspace{-1.5cm}-\sh\ln(\sh)\Big(
    8\pi^2+15\ln(\sh)-5\ln^2(\sh)+6\ln(\sh)\ln(1-\sh)\Big)\CR
  &\hspace{-1.5cm}+3\pi i\sh\ln(\sh)\Big(
    4\ln(1-\sh)+10-5\ln(\sh)\Big)
  -2\sh\pi^3 i
  +6\Big(1-3\sh\Big)\zeta(3)\Big]+O(\eps^2)\bigg\}
\end{align}
where 
\begin{eqnarray}
{\rm Li}_3(x)=\int_0^x dt\,\frac{{\rm Li}_2(t)}{t}
\end{eqnarray}
is the Trilogarithm.
\begin{align}
\begin{picture}(90,30)(0,14)
 \psfig{file=MI43.ps,width = 3cm}
\end{picture} 
&=\mu^{2(4-d)}\int\{d^dk\}\{d^dl\}
\frac{1}{{\cal P}_2{\cal P}_3{\cal P}_6{\cal P}_7}\CR
&=
-\left(\frac{m_b^2}{\mu^2}\right)^{-2\eps}\bigg\{
\frac{1}{2\eps^2}
+\frac{1}{2\eps}\Big(5-2\ln(\sh)+2\pi i\Big)\CR
&\hspace{-3cm}+{\rm Li}_2(\sh)-5\ln(\sh)+\frac{1}{2}\ln^2(\sh)
 +\ln(\sh)\ln(1-\sh)+\frac{19}{2}-\frac{4}{3}\pi^2
 -2\pi i\ln(\sh)+5\pi i+O(\eps)\bigg\}
\end{align}
\begin{align}
\begin{picture}(90,30)(0,14)
 \psfig{file=MI45.ps,width = 3cm}
\end{picture} 
&=\mu^{2(4-d)}\int\{d^dk\}\{d^dl\}
\frac{1}{{\cal P}_2{\cal P}_6{\cal P}_7{\cal P}_9}\CR
&\hspace{-1cm}=
-\left(\frac{m_b^2}{\mu^2}\right)^{-2\eps}\bigg\{
\frac{1}{2\eps^2}
+\frac{1}{2\eps}\Big(5-2\ln(\sh)+2\pi i\Big)
-5\ln(\sh)+\frac{19}{2}-\pi^2+5\pi i\CR
&\hspace{-3cm}+\frac{1}{6(1-\sh)}\Big[
  6\,{\rm Li}_3(\sh)
  -6\Big(\ln(\sh)-\pi i\Big){\rm Li}_2(\sh)
  -3\ln^2(\sh)\Big(\ln(1-\sh)+2\sh-1\Big)\CR
  &\hspace{-1cm}+\ln(\sh)\Big(\pi^2+6\pi i\ln(1-\sh)+12\pi i\sh-6\pi i\Big)
  -\pi^3 i-6\zeta(3)\Big]+O(\eps)\bigg\}
\end{align}
\begin{align}
\begin{picture}(90,30)(0,14)
 \psfig{file=MI452.ps,width = 3cm}
\end{picture} 
&=\mu^{2(4-d)}\int\{d^dk\}\{d^dl\}
\frac{1}{{\cal P}_2({\cal P}_6)^2{\cal P}_7{\cal P}_9}\CR
&\hspace{-1cm}=
-\frac{1}{m_b^2}\left(\frac{m_b^2}{\mu^2}\right)^{-2\eps}\bigg\{
\frac{1}{2(1-\sh)\eps}\ln(\sh)\Big(\ln(\sh)-2\pi i\Big)
-\frac{1}{6(1-\sh)}\Big[
  18\,{\rm Li}_3(\sh)\CR
  &\hspace{-2cm}-18\Big(\ln(\sh)-\pi i\Big){\rm Li}_2(\sh)
  +\ln(\sh)\Big(5\ln^2(\sh)-9\ln(\sh)\ln(1-\sh)
  -7\pi^2\CR
  &\hspace{-1cm}-15\pi i\ln(\sh)+18\pi i\ln(1-\sh)\Big)
  -3\pi^3 i-18\zeta(3)\Big]+O(\eps)\bigg\}
\end{align}
\begin{align}
\begin{picture}(90,30)(0,14)
 \psfig{file=MI44.ps,width = 3cm}
\end{picture} 
&=\mu^{2(4-d)}\int\{d^dk\}\{d^dl\}
\frac{1}{{\cal P}_2{\cal P}_5{\cal P}_6{\cal P}_9}\CR
&=
-\left(\frac{m_b^2}{\mu^2}\right)^{-2\eps}\bigg\{
\frac{1}{2\eps^2}
-\frac{1}{2\eps}\Big(
  4\sqrt{z-1}\,{\rm arccot}(\sqrt{z-1})-5\Big)\CR
&\hspace{-3cm}-\frac{\sqrt{z-1}}{12}\Big[i\Big(
  24\,{\rm Li}_2(x_2/x_1)-12\,{\rm Li}_2(-x_2/x_1)
  -15\ln^2(x_2/x_1)+12\ln^2(x_2)\CR
  &\hspace{-1cm}-12\ln(x_1)\ln(x_2)
  +24\ln(x_1)\ln(x_4)-24\ln(x_2)\ln(x_4)-5\pi^2\Big)\CR
  &\hspace{-1cm}+6\pi\Big(\ln(x_1)+\ln(x_2)-2\ln(z-1)\Big)
  +120\,{\rm arccot}(\sqrt{z-1})\Big]\CR
  &\hspace{-3cm}+\frac{114}{12}
+\frac{1}{12(1-\sh)}\Big[
  \Big(3-\sh\Big)\Big(9\ln^2(x_2/x_1)\CR
     &\hspace{.5cm}-36\pi\,{\rm arccot}(\sqrt{z-1})\Big)
  +\pi^2\Big(33-13\sh\Big)\Big]+O(\eps)\bigg\}
\end{align}
\begin{align}
\begin{picture}(90,30)(0,14)
 \psfig{file=MI442.ps,width = 3cm}
\end{picture} 
&=\mu^{2(4-d)}\int\{d^dk\}\{d^dl\}
\frac{1}{({\cal P}_2)^2{\cal P}_5{\cal P}_6{\cal P}_9}\CR
&=
\frac{1}{6a(1-\sh)\eps}
  \Big(36\,{\rm arccot}^2(\sqrt{z-1})-\pi^2\Big)+O(\eps^0)
\end{align}

\section{Unrenormalized matrix elements
         $\langle d\ell^+\ell^-|{\cal O}_{1,2}^u|b\rangle$}

\label{appUnrenFF}

We present the results of the matrix--elements as in section~\ref{secRenFF}. After splitting up into the different diagrams and the different color structures that can arise, one gets
\begin{eqnarray}
F_{1,u}^{{\rm fig.}(7)} &=& \left(\frac{m_b^2}{\mu^2}\right)^{-2\eps}
  A^{\rm fig.}(s),\\
F_{2,u}^{{\rm fig.}(7)} &=& -6 \left(\frac{m_b^2}{\mu^2}\right)^{-2\eps}
  A^{\rm fig.}(s),\\
F_{1,u}^{{\rm fig.}(9)} &=& \left(\frac{m_b^2}{\mu^2}\right)^{-2\eps}
  \Big(B^{\rm fig.}(s) + 4 C^{\rm fig.}(s)\Big),\\
F_{2,u}^{{\rm fig.}(9)} &=& \left(\frac{m_b^2}{\mu^2}\right)^{-2\eps}
  \Big(-6 B^{\rm fig.}(s) + 3 C^{\rm fig.}(s)\Big),
\end{eqnarray}
where the functions $A^{\rm fig.}(s)$, $B^{\rm fig.}(s)$ and $C^{\rm fig.}(s)$ for the individual group of diagrams are given in the subsequent paragraphs up to the zeroth order in $\eps$.

\paragraph{Diagrams shown in fig.~\ref{AbballeDiag}a}

\begin{eqnarray}
A^{\ref{AbballeDiag}a}(s)&=&
\frac{2}{27\eps}
+\frac{5}{9}+\frac{4}{27}\,\pi i\CR
&&\hspace{-1cm}+\frac{2\sh}{81(1-\sh)}\,\Big[
6\,{\rm Li}_2(\sh)-3\ln^2(\sh)+6\ln(\sh)\Big(1+\ln(1-\sh)\Big)
-\pi^2\Big]\\
B^{\ref{AbballeDiag}a}(s)&=&
\frac{4}{27\eps^2}
-\frac{2}{27\eps}\Big(4\ln(\sh)-5-4\pi i\Big)
+\frac{1}{9}+\frac{20}{27}\,\pi i-\frac{16}{27}\,\pi i \ln(\sh)\CR
&&\hspace{-1cm}+\frac{2}{81(1-\sh)}\,\Big[
  12\Big(1-2\sh\Big)\Big({\rm Li}_2(\sh)+\ln(1-\sh)\ln(\sh)\Big)\CR
  &&\hspace{1.5cm}+6\ln(\sh)\Big(\ln(\sh)-9+5\sh\Big)
  -\pi^2\Big(15-17\sh\Big)\Big]
\end{eqnarray}

\paragraph{Diagrams shown in fig.~\ref{AbballeDiag}b}

\begin{eqnarray}
A^{\ref{AbballeDiag}b}(s)&=&
\frac{4}{27\eps}
+\frac{2}{27(1-\sh)^2}\,\Big[
  7-15\sh+8\sh^2
  -\pi i \sh\Big(1-3\sh\Big)\Big]\CR
&&\hspace{-1cm}+\frac{2\sh}{27(1-\sh)^3}\,\Big[
  \Big(1-4\sh+3\sh^2-2\pi i(1-2\sh)\Big)\ln(\sh)
  +(1-2\sh)\ln^2(\sh)\Big]\\
B^{\ref{AbballeDiag}b}(s)&=&
\frac{4}{27\eps^2}
-\frac{2}{27\eps}\Big(4\ln(\sh)-5-4\pi i\Big)
-\frac{2}{9}\pi^2
+\frac{1}{27(1-\sh)^2}\,\Big[
  7-10\sh+3\sh^2\CR
  &&\hspace{-1cm}+4\pi i(6-13\sh+5\sh^2)\Big]
-\frac{4}{27(1-\sh)^3}\,\ln(\sh)\Big[
  6-19\sh+18\sh^2-5\sh^3\CR
  &&\hspace{-1cm}+2\pi i(1-4\sh+6\sh^2-2\sh^3)
  -(1-4\sh+6\sh^2-2\sh^3)\ln(\sh)\Big]
\end{eqnarray}

\paragraph{Diagrams shown in fig.~\ref{AbballeDiag}c}

\begin{eqnarray}
C^{\ref{AbballeDiag}c}(s)&=&
\frac{8}{27\eps}
-\frac{4}{81}\,\Big(12\ln(\sh)-49+48\zeta(3)-12 \pi i\Big)
\end{eqnarray}

\paragraph{Diagrams shown in fig.~\ref{AbballeDiag}d}

\begin{eqnarray}
A^{\ref{AbballeDiag}d}(s)&=&
\frac{2}{243\eps}
+\frac{1}{729}\,\Big(37+12\pi i\Big)
+\frac{4\sh}{243(1-\sh)}\,\ln(\sh)\\
B^{\ref{AbballeDiag}d}(s)&=&
-\frac{4}{243\eps^2}
+\frac{2}{729\eps}\Big(12\ln(\sh)-19-12\pi i\Big)
-\frac{1}{2187}\,\Big(72\,{\rm Li}_2(\sh)+72\ln(1-\sh)\ln(\sh)\CR
  &&\hspace{-1cm}+36\ln^2(\sh)+463-90\pi^2-12\pi i(12\ln(\sh)-19)\Big)
+\frac{4}{729(1-\sh)}\,\ln(\sh)(13-19\sh)
\end{eqnarray}

\paragraph{Diagrams shown in fig.~\ref{AbballeDiag}e}

\begin{eqnarray}
A^{\ref{AbballeDiag}e}(s)&=&
-\frac{10}{243\eps}
+\frac{4\sqrt{z-1}}{243(1-\sh)^3}\,
  \Big(10-27\sh+33\sh^2-10\sh^3\Big){\rm arccot}(\sqrt{z-1})\CR
&&\hspace{-1cm}+\frac{1}{729(1-\sh)^4}\, \Big(
  -107+452\sh-696\sh^2+464\sh^3-113\sh^4-6\pi^2\sh\CR
  &&+216\sh \,{\rm arccot}^2(\sqrt{z-1})\Big)\\
B^{\ref{AbballeDiag}e}(s)&=&
-\frac{4}{243\eps^2}
+\frac{2}{729\sh\eps}\Big(
  24\sqrt{z-1}(2+\sh){\rm arccot}(\sqrt{z-1})
  -48-19\sh\Big)\CR
&&\hspace{-1cm}+\frac{2\sqrt{z-1}(2+\sh)}{729\sh}\Big[\,
  6\pi\Big(\ln(x_1)+\ln(x_2)-2\ln(z-1)\Big)\CR
  &&+i\Big(24{\rm Li}_2(x_2/x_1)-12{\rm Li}_2(-x_2/x_1)-15\ln^2(x_2/x_1)
    +12\ln^2(x_2)\CR
    &&\hspace{1cm}-12\ln(x_1)\ln(x_2)+24\ln(x_1)\ln(x_4)
    -24\ln(x_2)\ln(x_4)-5\pi^2\Big)\Big]\CR
&&\hspace{-1cm}+\frac{2}{81\sh(1-\sh)}\,
  \Big(6+\sh-\sh^2\Big)\Big(4\pi{\rm arccot}(\sqrt{z-1})-\ln^2(x_2/x_1)\Big)\CR
&&\hspace{-1cm}-\frac{1}{2187\sh(1-\sh)^2}\,\Big(
  1056-1541\sh+130\sh^2+463\sh^3\Big)\CR
&&\hspace{-1cm}+\frac{8\sqrt{z-1}}{729\sh(1-\sh)^3}\,
  \Big(32-65\sh-6\sh^2+40\sh^3-19\sh^4\Big){\rm arccot}(\sqrt{z-1})\CR
&&\hspace{-1cm}+\frac{4}{729\sh(1-\sh)^4}\, \Big[
  36\Big(3-8\sh+2\sh^2\Big){\rm arccot}^2(\sqrt{z-1})\CR
  &&-\pi^2\Big(27-67\sh+28\sh^2+36\sh^3-34\sh^4+7\sh^5\Big)\Big]
\end{eqnarray}

\section{One--loop matrix elements to O($\eps$)}

\label{appOneloop}

The one--loop matrix elements
\begin{eqnarray}
\langle d\ell^+\ell^-|{\cal O}_i^u|b\rangle_{\rm one-loop}=
F_{i,u}^{{\rm 1-loop}(7)}\langle {\cal \tilde O}_7\rangle_{\rm tree}
+F_{i,u}^{{\rm 1-loop}(9)}\langle {\cal \tilde O}_9\rangle_{\rm tree}
\end{eqnarray}
are needed to O($\eps$) for the operators ${\cal O}_{1,2,4,11,12}$. The results are
\begin{eqnarray}
F_{1,u}^{{\rm 1-loop}(9)} &=&\left(\frac{s}{\mu^2}\right)^{-\eps}\bigg\{
\frac{16}{27\eps}
+\frac{16}{81}\,\Big(2+3\pi i\Big)
+\frac{4\eps}{243}\,\Big(52-21\pi^2+24\pi i\Big)\bigg\}\\
F_{2,u}^{{\rm 1-loop}(9)} &=&
\frac{3}{4}\,F_{1,u}^{{\rm 1-loop}(9)}\\
F_{4,u}^{{\rm 1-loop}(7)} &=&\left(\frac{m_b^2}{\mu^2}\right)^{-\eps}\bigg\{
-\frac{4}{9}
+\frac{8\eps}{9}\,\Big(
\sqrt{z-1}\,{\rm arccot}(\sqrt{z-1})-1\Big)\bigg\}\\
F_{4,u}^{{\rm 1-loop}(9)}&=&\left(\frac{m_b^2}{\mu^2}\right)^{-\eps}\bigg\{
-\frac{16}{27\eps}\CR
&&\hspace{-1cm}+\frac{8}{81\sh}\,\Big[
6(2+\sh)\sqrt{z-1}\,{\rm arccot}(\sqrt{z-1})
+3\sh\ln(\sh)-12-4\sh-3\pi i\sh\Big]\CR
&&\hspace{-1cm}
+\frac{\eps(2+\sh)\sqrt{z-1}}{81\sh}\,\Big[
i\Big(48\,{\rm Li}_2(x_2/x_1)
+24\ln^2(x_1)-12\ln^2(x_2)+12\ln^2(x_4)\CR
&&+24\ln(x_1)\ln(x_3)-48\ln(x_2)\ln(x_4)-3\ln^2(z-1)
-5\pi^2\Big)\CR
&&+6\pi\Big(6\ln(x_1)+2\ln(x_2)+2\ln(x_3)
+2\ln(x_4)-\ln(z-1)\Big)\Big]\CR
&&\hspace{-1cm}+\frac{32\eps(5+\sh)\sqrt{z-1}}{81\sh}\,{\rm arccot}(\sqrt{z-1})\CR
&&\hspace{-1cm}-\frac{4\eps}{243\sh}\Big[
120+(52+12\pi i-9\pi^2)\sh
-6(2+3\pi i)\sh\ln(\sh)+9\sh\ln^2(\sh)\Big]\bigg\}\\
F_{11,u}^{{\rm 1-loop}(9)}&=&\left(\frac{s}{\mu^2}\right)^{-\eps}\bigg\{
-\frac{64}{27}
-\frac{64\eps}{81}\Big(5+3\pi i\Big)\bigg\}\\
F_{12,u}^{{\rm 1-loop}(9)}&=&
\frac{3}{4}\,F_{11,u}^{{\rm 1-loop}(9)}
\end{eqnarray}

\end{appendix}

\end{document}